\documentclass[a4paper]{article}
\usepackage{amsmath, amssymb, graphicx, algorithm, algpseudocode, caption, subcaption}
\usepackage[table]{xcolor}
\usepackage[backend=biber,urldate=iso,date=iso,style=numeric]{biblatex}
\newtheorem{lemma}{Lemma}

\title{Computational bounds for the 2048 game}
\author{Alexey Slizkov\thanks{elexunix@gmail.com; National Research University Higher School of Economics, Moscow, Russian Federation}}
\date{}

\begin{document}
\maketitle

\begin{abstract}
2048 is a single player video game, played by millions mostly on mobile devices. We prove rigorously for the first time that there is an algorithm with winning probability at least 0.99969, and that there is a strategy for achieving the 256 tile guaranteed (with probability 1).
\end{abstract}

\section{Introduction}

2048 is a game developed by Gabriele Cirulli, an Italian web developer, and hosted on GitHub in March 2014. This game is a sliding tile puzzle designed for solo play, where your goal is to merge numbered tiles on a grid until you reach the tile with the number 2048. The game does not stop once this goal is achieved, allowing you to continue playing and attempting to create larger numbered tiles \cite{wikipedia}. People have long been exploring theoretical and practical aspects of the game. For example, \cite{langerman} and \cite{mehta} prove some hardness results (NP-hardness and PSPACE-hardness). On the other hand, there are plenty of works concerning heuristics and applicability of machine learning strategies for the game, for example, \cite{rodgers-levine} applies Monte-Carlo Tree Search, while \cite{boris-goran}, \cite{matsuzaki} discuss other machine learning approaches.

\begin{figure}
  \centering
  \includegraphics[width=300pt]{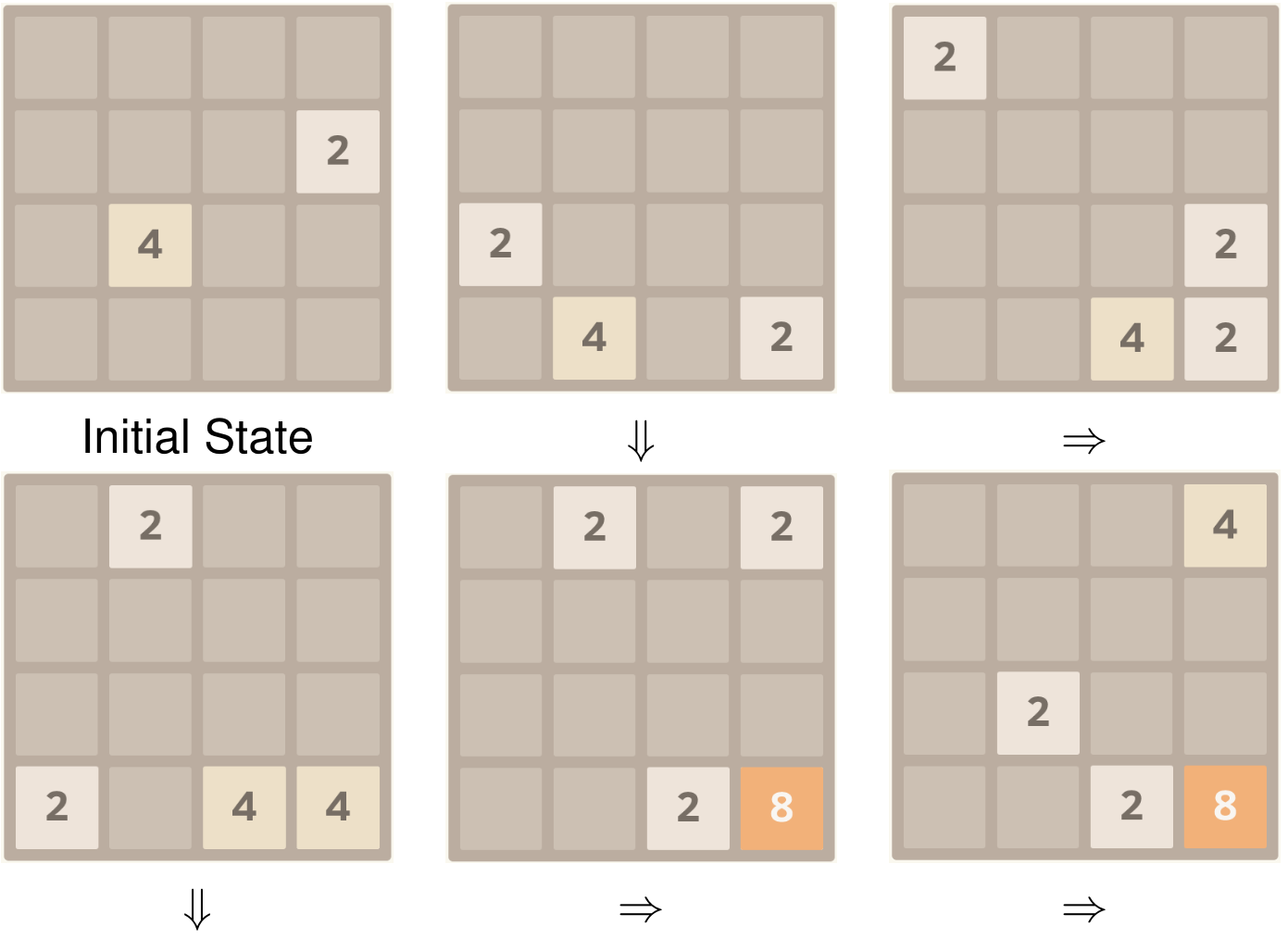}
  \caption{First six moves of a game of 2048. Source: \cite{mehta}}
\end{figure}

Remember that tiles appear in random free cells with equal probability, with 90\% probability of 2 and 10\% probability of 4, this is where the probabilities come from.

One of the most natural questions is what is the optimal winning probability in this game. AI-based algorithms have shown that it is probably very high, but no proof of this existed before. We show for the first time that the winning probability is at least 0.99969486 by selecting and traversing a relatively small set of positions in which the player can stay and win with high probability, using only integers in our computation.

The natural game-theoretic question is whether there is a strategy that reaches 2048 always (that is, the aforementioned probability is exactly 1), or there is a strategy that can always prevent the player from achieving 2048. We explore the game on smaller field sizes, and prove the 256 lower bound for the maximum achievable tile, using a three months-long computation on a machine with over 500GB or RAM and 80 to 96 cores. We also show several other similar results regarding guaranteed reachability of different tiles on fields of different sizes, this time using a machine with an RTX 4090 GPU, 64GB RAM and 2x 3GB/s SSDs for up to several days per computation.

\section{Guaranteed reachability}

Let us fix a field size $m \times n$ and a goal tile $T$. Let's consider the graph $G_{m,n,T}$ of all $m \times n$ positions with all tiles less than $T$. For the questions of guaranteed reachability and winning probability only such positions have to be considered. Note that there are $2(\log_2 T)^{m\times n}$ such positions (each cell can be either empty or have tile $2^1$, $2^2$, \dots, or $2^{\log_2T-1}$, and there may be either player's turn or computer's turn). For $m = n = 4$ and $T = 256$ this gives us $2 \cdot 8^{16} \approx 5.63 \cdot 10^{14}$ positions, a feasible number, but this obviously doesn't fit in RAM or a drive. The basic idea is to traverse this graph in some order and mark each vertex as winning or losing (for the goal of achieving $T$); the terminal positions are those where either two $T / 2$ tiles can be merged or no move can be played. The answer, who wins the game, lies in the data for initial fields\footnote{The original game starts with two random tiles on the field, but for definiteness and simplicity (to eliminate the dependance on these two tiles) we consider the game starting with empty field and computer's move. However, for our result of guaranteed reachability of 256 we proved that it doesn't depend on the starting configuration as long as there are two tiles and it is the player's turn, in all $4 \cdot {16 \choose 2} = 240$ cases it turned out that the game is winnable}.

\begin{figure}
  \centering
  \includegraphics[width=360pt]{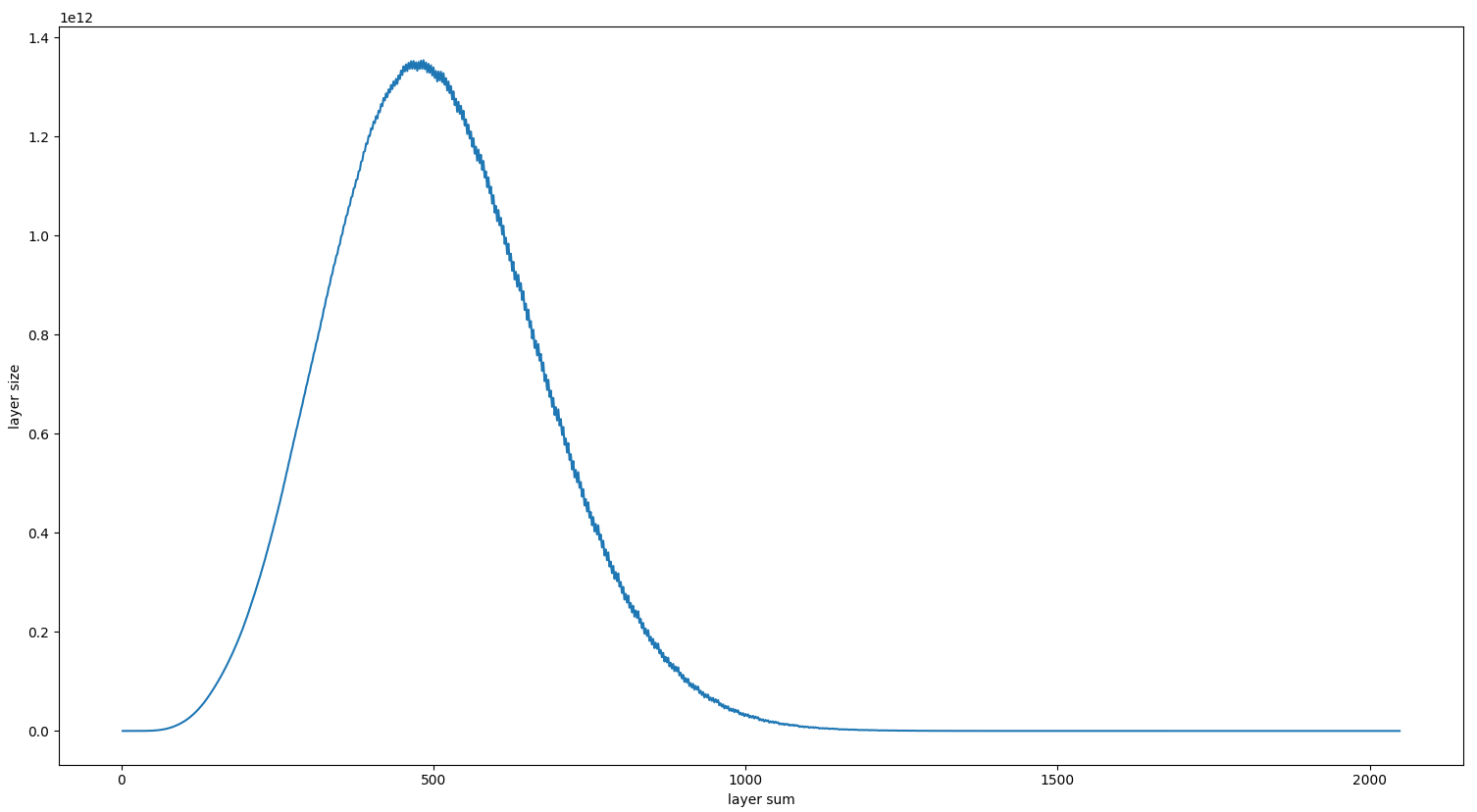}
  \caption{Sizes of layers into which we split the game graph $G_{m,n,T}$}
  \label{fig:layer-sizes}
\end{figure}

To solve the problem with required memory, we have split the game graph into \textit{layers} by sum: the sum of a position is the sum of all tile numbers on it, 0 for empty cells. Notice that the player's swipe doesn't change the sum even if some tiles merge, and the computer's move (adding the new tile) increases the sum by exactly 2 or 4. If we split the graph into layers of fixed sum and turn as follows and traverse it in this order, we will be able to store only three of these layers at once: [sum $mnT / 2$, player's turn], [sum $mnT / 2 - 2$, computer's turn], [sum $mnT / 2 - 2$, player's turn], [sum $mnT / 2 - 4$, computer's turn], \dots, [sum 4, computer's turn], [sum 4, player's turn], [sum 2, computer's turn], [sum 2, player's turn], [sum 0, computer's turn]. But how large are these layers?
We wrote an additional program to calculate the number of positions in these layers, they turned out to be as shown on figure~\ref{fig:layer-sizes}, the maximum layer size being $1353817378016 \approx 1.35 \cdot 10^{12}$. We decided to store one bit per position, so we needed 508GB or RAM. The parallelization into up to 96 threads was performed by splitting the layer into chunks of size $2^{14}$ positions and assigning each chunk to a thread pseudo-randomly using hash of the chunk index. The straightforward division without hash turned out to split the work unequally between the threads, the load was not balanced. But at this point another question comes into play: we need an effective way to index these layers stored as a large arrays of bits. We need to be able to solve two types of problems: fast indexation, get the position by its number inside a layer, and get the index of a given position.

Both problems are solved by means of dynamic programming using a precomputed array of size $(mn + 1) \times (mnT / 2 + 1)$, which stores, for each number of cells and their sum of tile numbers, the number of possible ways to fill this number of cells by tiles and empty cells so that the sum is as requires. For example, the value of the $(5, 216)$-th cell of the array is the number of quintuples of five tiles or empty cells with the total sum of 216. This array is precomputed once at the start of the program in order of increasing number of cells, starting from 0 cells, where the number of ways is 1 for sum 0 and 0 for all positive sums. The process is described in detail in algorithm 1. After we have computed this auxiliary table, call it $I$ (as it is used for indexation subproblems), we can find the index of a given position inside layer as follows: Let $a_1$, \dots, $a_{mn}$ be its cell tile numbers, 0 for empty cells, in order. Then the index is
$$ \sum\limits_{i=1}^{mn} \begin{cases} 0, \text{ if } a_i = 0,\\ I\bigg[mn - i\bigg]\bigg[\sum\limits_{j=i+1}^{mn} a_j\bigg] + \sum\limits_{k=1}^{\log_2a_i-1} I\bigg[mn - i\bigg]\bigg[\sum\limits_{j=i+1}^{mn} a_j - 2^k\bigg], \text{ otherwise}. \end{cases} $$
The inverse problem, to find the position by its index, is solved by algorithm~\ref{alg:pos-by-idx}.

\begin{algorithm}
  \begin{algorithmic}
    \For{$s \gets 0$, 2, \dots, $mnT / 2$}
      \If{$s = 0$}
        \State $I[0][s] \gets 1$
      \Else
        \State $I[0][s] \gets 0$
      \EndIf
    \EndFor
    \For{$c \gets 1$, \dots, $mn$}
      \For{$s \gets 0$, 2, \dots, $mnT / 2$}
        \State $I[c][s] \gets I[c - 1][s]$
        \State $t \gets 2$
        \While{$t \leq s$}
          \State $I[c][s] \gets I[c][s] + I[c - 1][s - t]$
          \State $t \gets 2 \cdot t$
        \EndWhile
      \EndFor
    \EndFor
  \end{algorithmic}
  \caption{Filling the $I$ auxiliary table}
  \label{alg:fill-dp}
\end{algorithm}

\begin{algorithm}[!h]
  \begin{algorithmic}
    \State $r \gets \text{index}$
    \State $s \gets \text{sum}$
    \For{$i \gets 1$, \dots, $mn$}
      \If{$r < I[mn - i][s]$}
        \State $a_i \gets 0$
      \Else
        \State $r \gets r - I[mn - i][s]$
        \State $t \gets 2$
        \While{$r < T[mn - i][s - t]$}
          \State $r \gets r - I[mn - i][s - t]$
          \State $t \gets 2 \cdot t$
        \EndWhile
        \State $a_i \gets t$
        \State $s \gets s - t$
      \EndIf
    \EndFor
    \Comment{at this point, $r = 0$ and $s = 0$}
  \end{algorithmic}
  \caption{Find the position by its sum and index inside layer with that sum}
  \label{alg:pos-by-idx}
\end{algorithm}

The computation proved that 256 is reachable. The author has saved all the binary arrays of reachability and compressed them to 9TB, which are currently stored in the NRU HSE Faculty of Computer Science distributed storage.

We have performed similar computations, some using the CUDA technology to program on GPU to speed up the computation, and our main guaranteed reachability results can be summarized in table~\ref{fig:gr-table}. Additional results can be found in the Appendix.

\begin{figure}
  \centering
  \scalebox{1.7}{
      \begin{tabular}{ | c | c c c c c c | }
      \hline
      $m \backslash n$ & 1 & 2 & 3 & 4 & 5 & 6 \\
      \hline
      1 & 2 & 4 & 4 & 4 & 4 & 4 \\
      2 & 4 & 8 & 16 & 32 & 32 & 64 \\
      3 & 4 & 16 & 32 & 64 & 128 & \\
      4 & 4 & 32 & 64  & ? & & \\
      5 & 4 & 32 & 128 & & & \\
      6 & 4 & 64 & & & & \\
      \hline
    \end{tabular}
  }
  \caption{The maximum tiles guaranteed achievable on different field sizes}
  \label{fig:gr-table}
\end{figure}

However, this table indicates that the maximum achievable tile on the original field size is probably much smaller than 2048, and therefore, there exists a strategy of tile generation that never allows the player to attain 2048. We have some ideas on how we can organize the computation for tile 512 without terabytes of RAM and years of computing time, although the graph contains over 3.7 quadrillion nodes.

\section{Lower bounds on probabilities}

The computation for lower bound on probabilities relies on the same ideas about layers of constant sum and effective indexing inside them. But this time, we have to store much more information per position than one bit, this time we have to store a lower bound on probability, a real number, in each node. We have computed some results in floating point numbers approximately (see Appendix), but we have also made our program prove lower bounds using only integer arithmetic, always rounding to the lower, and it turned out that the type \verb|uint16_t| was not enough to store probability (times $2^{16}$), it gave very low bounds, but \verb|uint32_t| suited our needs (and \verb|uint64_t| gave very little new precision, so we didn't use it).

For our main result (probability 0.99969486 on the original field $4 \times 4$ for the original tile 2048), we have severely restricted the set of allowed positions to make the computation feasible, and in fact it was completed in just one day and a half, mostly bottlenecking in disk bandwidth (3GB/s). We have used SSD disks, and our total TeraBytes Written (TBW) for all computations was approximately 400TB, and about 1.2PB was read. On the machine with the GPU where this computation was performed, there was only 64GB of RAM, and the maximum layer size was bigger. Actually, the computation used the following hierarchical structure of available resources: 24GB was the very fast, 1TB/s GDDR6 GPU memory, where the most frequently accessed data was stored; then, 64GB was the host RAM, our configuration had a bandwidth of 41GB/s, and the management of data transferring between the host RAM and the GPU memory was completely placed under management of Nvidia CUDA Compiler and drivers via calls like \verb|cudaMallocManaged| in the CUDA API; next, a fast Solid State Drive with read/write speed of around 3GB/s, stored the whole three layers we operated over. We have batched each layer into 32GB batches so that two such batches fit into memory. The algorithm~\ref{alg:chunked} describes the process in more detail.

\begin{algorithm}
  \centering
  \begin{algorithmic}
    \For{$s \gets \text{maxsum}$, $\text{maxsum} - 2$, \dots, 0}
      \For{each batch $B_0$ in computer's layer with sum $s$}
        \State Initialize batch $B_0$ with zero probabilities
        \For{each batch $B_1$ in player's layer with sum $s + 2$}
          \State Read batch $B_1$ from disk
          \State Add probabilities implied by batch $B_1$ data to batch $B_0$
        \EndFor
        \For{each batch $B_1$ in player's layer with sum $s + 4$}
          \State Read batch $B_1$ from disk
          \State Add probabilities implied by batch $B_1$ data to batch $B_0$
        \EndFor
        \State In batch $B_0$, divide the accumulated sums of probabilities by the number of free cells
        \State Write batch $B_0$ to disk
      \EndFor
      \State Remove player's layer with sum $s + 4$ from disk if present
      \For{each batch $B_0$ in player's layer with sum $s$}
        \State Initialize batch $B_0$ with zero probabilities
        \State In batch $B_0$, replace terminal winning nodes' probabilities to one
        \For{each batch $B_1$ in computer's layer with sum $s$}
          \State In batch $B_0$, update probabilities using batch $B_1$
        \EndFor
      \EndFor
    \EndFor
  \end{algorithmic}
  \caption{Details of processing layers that don't fit into RAM}
  \label{alg:chunked}
\end{algorithm}

\begin{figure}
  \centering
  \begin{subfigure}{.5\textwidth}
    \centering
    \includegraphics[width=150pt]{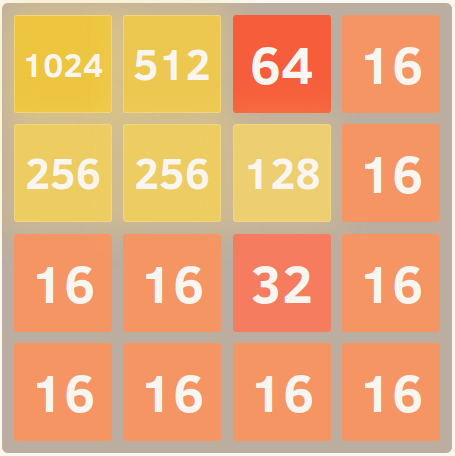}
    \caption{Maximum allowed tiles for each cell}
    \label{fig:max-tiles}
  \end{subfigure}%
  \begin{subfigure}{.5\textwidth}
    \centering
    \includegraphics[width=150pt]{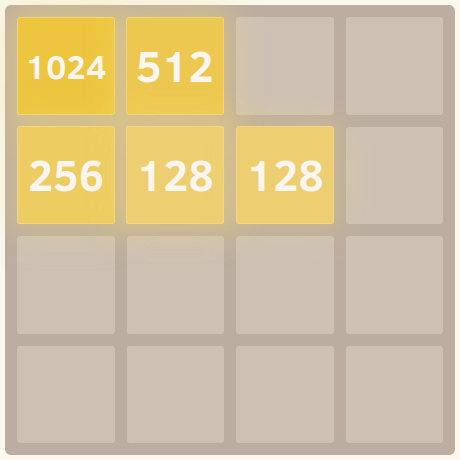}
    \caption{Required tiles winning configuration}
    \label{fig:tiles-req}
  \end{subfigure}
  \caption{Maximum and required tiles in the computation}
\end{figure}

The huge reduction in the number of positions to be considered was achieved by limiting the tiles on each cell with the values on figure~\ref{fig:max-tiles}. This reduced the number of states to be considered from $2 \cdot 11^{16} \approx 9.19 \cdot 10^{16}$ to $11 \cdot 10 \cdot 9^2 \cdot 8 \cdot 7 \cdot 6 \cdot 5^9 \approx 1.17 \cdot 10^{13}$, almost 8000-fold. But one may notice that it is impossible to form two 1024s near each other using these limitations. Indeed, we set the goal to have the large tiles as on figure~\ref{fig:tiles-req} on their places instead. One can prove that this is enough to get the 2048 tile (see Lemma~\ref{lemma:the-lemma}), and it turns out that these limitations on how we get the 2048 tile still allow us to have a winning probability lower bound of 0.99969486.

\begin{lemma} \label{lemma:the-lemma}
  Whatever the other tiles are, the configuration with the five big tiles as in Figure~\ref{fig:tiles-req} is winning.
\end{lemma}

\section{Future work}

For the traversal of even larger graphs, those like the graph $G_{4,4,512}$ with approximately $3.71 \cdot 10^{15}$ states, we need to split the layers even further, as the batch approach becomes infeasible due to its quadratic complexity: at some point of the computation, the three layers' total size will be 1.7TB. And although we cannot reduct this amount of data by traversing a graph in another way, we can explore in finer details the graph edges between edges. The following idea seems to be promising: split the layers further into groups of fixed set of tiles, there will be 735471 groups in our case, with the maximum group size of only 163 billion positions (20GB).

And the method with which we proved the lower bound for the original winning probability can be used, with the aforementioned edition, to prove lower bounds for much larger tiles, for example, 32768 and 65536, which are of interest to some researchers of the problem: for example,~\cite{xue} reach 65536 around 3\% of the time, claiming to have significantly beaten the previous record of around 0.02\%.

\printbibliography[heading=bibintoc]

\appendix

\section{Additional computational results}

\subsection{Guaranteed reachability}

The following table summarizes guaranteed reachability results:\\

\begin{tabular}{ | c | c | c | c | }
  \hline
  Field size & Max allowed tiles during play & Goal & Result\\
  \hline\hline
  $2 \times 4$ & \begin{tabular}{ c c c c } 16 & 16 & 16 & 8\\ 16 & 16 & 16 & 8\\ \end{tabular} & 32 & \cellcolor{green!25} Winnable, minimal\\
  \hline
  $2 \times 4$ & \begin{tabular}{ c c c c } 32 & 32 & 32 & 32\\ 32 & 32 & 32 & 32\\ \end{tabular} & 64 & \cellcolor{red!25} Not winnable\\
  \hline\hline
  $3 \times 3$ & \begin{tabular}{ c c c } 16 & 16 & 16\\ 16 & 16 & 16\\ 8 & 8 & 8\\ \end{tabular} & 32 & \cellcolor{green!25} Winnable, minimal\\
  \hline
  $3 \times 3$ & \begin{tabular}{ c c c } 16 & 16 & 8\\ 16 & 16 & 8\\ 8 & 16 & 16\\ \end{tabular} & 32 & \cellcolor{green!25} Winnable, minimal\\
  \hline
  $3 \times 3$ & \begin{tabular}{ c c c } 16 & 8 & 16\\ 16 & 16 & 16\\ 16 & 8 & 16\\ \end{tabular} & 32 & \cellcolor{red!25} Not winnable, maximal\\
  \hline
  $3 \times 3$ & \begin{tabular}{ c c c } 16 & 16 & 16\\ 16 & 8 & 16\\ 16 & 16 & 16\\ \end{tabular} & 32 & \cellcolor{red!25} Not winnable, maximal\\
  \hline
  $3 \times 3$ & \begin{tabular}{ c c c } 32 & 32 & 32\\ 32 & 32 & 32\\ 32 & 32 & 32\\ \end{tabular} & 64 & \cellcolor{red!25} Not winnable\\
  \hline\hline
  $2 \times 5$ & \begin{tabular}{ c c c c c } 16 & 16 & 16 & 16 & 8\\ 16 & 16 & 16 & 8 & 8\\ \end{tabular} & 32 & \cellcolor{green!25} Winnable, minimal\\
  \hline
  $2 \times 5$ & \begin{tabular}{ c c c c c } 32 & 32 & 32 & 32 & 32\\ 32 & 32 & 32 & 32 & 32\\ \end{tabular} & 64 & \cellcolor{red!25} Not winnable\\
  \hline\hline
  $2 \times 6$ & \begin{tabular}{ c c c c c c } 16 & 32 & 32 & 32 & 32 & 16\\ 16 & 32 & 32 & 32 & 32 & 16\\ \end{tabular} & 64 & \cellcolor{green!25} Winnable, minimal\\
  \hline
  $2 \times 6$ & \begin{tabular}{ c c c c c c } 64 & 64 & 64 & 64 & 64 & 64\\ 64 & 64 & 64 & 64 & 64 & 64\\ \end{tabular} & 128 & \cellcolor{red!25} Not winnable\\
  \hline\hline
  $3 \times 4$ & \begin{tabular}{ c c c c } 32 & 32 & 32 & 16\\ 16 & 32 & 32 & 16\\ 16 & 16 & 16 & 16\\ \end{tabular} & 64 & \cellcolor{green!25} Winnable, minimal\\
  \hline
  $3 \times 4$ & \begin{tabular}{ c c c c } 16 & 32 & 32 & 16\\ 32 & 32 & 32 & 16\\ 16 & 16 & 16 & 8\\ \end{tabular} & 64 & \cellcolor{green!25} Winnable, minimal\\
  \hline
  $3 \times 4$ & \begin{tabular}{ c c c c } 32 & 32 & 16 & 32\\ 32 & 32 & 32 & 16\\ 32 & 16 & 16 & 32\\ \end{tabular} & 64 & \cellcolor{red!25} Not winnable, maximal\\
  \hline
  $3 \times 4$ & \begin{tabular}{ c c c c } 32 & 32 & 32 & 32\\ 32 & 32 & 16 & 32\\ 32 & 32 & 16 & 32\\ \end{tabular} & 64 & \cellcolor{red!25} Not winnable, maximal\\
  \hline
  $3 \times 4$ & \begin{tabular}{ c c c c } 16 & 32 & 32 & 16\\ 16 & 32 & 32 & 16\\ 32 & 16 & 16 & 32\\ \end{tabular} & 64 & \cellcolor{red!25} Not winnable, maximal\\
  \hline
\end{tabular}

\newpage

\begin{tabular}{ | c | c | c | c | }
  \hline
  Field size & Max allowed tiles during play & Goal & Result\\
  \hline\hline
  $3 \times 4$ & \begin{tabular}{ c c c c } 64 & 64 & 64 & 64\\ 64 & 64 & 64 & 64\\ 64 & 64 & 64 & 64\\ \end{tabular} & 128& \cellcolor{red!25} Not winnable\\
  \hline\hline
  $3 \times 5$ & \begin{tabular}{ c c c c c } 64 & 64 & 64 & 32 & 32\\ 64 & 64 & 64 & 32 & 32\\ 32 & 32 & 32 & 32 & 32\\ \end{tabular} & 128 & \cellcolor{green!25} Winnable\\
  \hline
  $3 \times 5$ & \begin{tabular}{ c c c c c } 64 & 64 & 32 & 32 & 32\\ 64 & 64 & 64 & 32 & 32\\ 32 & 32 & 32 & 32 & 32\\ \end{tabular} & 128 & \cellcolor{red!25} Not winnable\\
  \hline
  $3 \times 5$ & \begin{tabular}{ c c c c c } 64 & 64 & 64 & 32 & 32\\ 64 & 64 & 32 & 32 & 32\\ 32 & 32 & 32 & 32 & 32\\ \end{tabular} & 128 & \cellcolor{red!25} Not winnable\\
  \hline
  $3 \times 5$ & \begin{tabular}{ c c c c c } 32 & 64 & 64 & 64 & 16\\ 32 & 64 & 64 & 64 & 32\\ 32 & 32 & 32 & 32 & 32\\ \end{tabular} & 128 & \cellcolor{green!25} Winnable\\
  \hline
  $3 \times 5$ & \begin{tabular}{ c c c c c } 64 & 32 & 32 & 32 & 32\\ 32 & 64 & 64 & 64 & 32\\ 32 & 32 & 32 & 32 & 32\\ \end{tabular} & 128 & \cellcolor{red!25} Not winnable\\
  \hline
  $3 \times 5$ & \begin{tabular}{ c c c c c } 64 & 64 & 64 & 64 & 32\\ 64 & 64 & 64 & 32 & 32\\ 32 & 32 & 32 & 32 & 32\\ \end{tabular} & 128 & \cellcolor{green!25} Winnable\\
  \hline
  $3 \times 5$ & \begin{tabular}{ c c c c c } 64 & 64 & 64 & 32 & 32\\ 64 & 64 & 64 & 32 & 32\\ 32 & 32 & 32 & 32 & 32\\ \end{tabular} & 128 & \cellcolor{red!25} Not winnable\\
  \hline
  $3 \times 5$ & \begin{tabular}{ c c c c c } 64 & 64 & 64 & 32 & 32\\ 64 & 64 & 64 & 32 & 32\\ 64 & 64 & 64 & 32 & 32\\ \end{tabular} & 128 & \cellcolor{green!25} Winnable\\
  \hline
  $3 \times 5$ & \begin{tabular}{ c c c c c } 64 & 64 & 64 & 32 & 32\\ 64 & 64 & 64 & 32 & 32\\ 64 & 64 & 32 & 32 & 32\\ \end{tabular} & 128 & \cellcolor{red!25} Not winnable\\
  \hline
  $3 \times 5$ & \begin{tabular}{ c c c c c } 128 & 128 & 128 & 128 & 128\\ 128 & 128 & 128 & 128 & 128\\ 128 & 128 & 128 & 128 & 128\\ \end{tabular} & 256 & \cellcolor{red!25} Not winnable\\
  \hline\hline
  $4 \times 4$ & \begin{tabular}{ c c c c } 32 & 64 & 64 & 32\\ 32 & 64 & 64 & 32\\ 32 & 64 & 32 & 32\\ 32 & 32 & 32 & 32\\ \end{tabular} & 128 & \cellcolor{green!25} Winnable\\
  \hline
  $4 \times 4$ & \begin{tabular}{ c c c c } 32 & 64 & 32 & 32\\ 32 & 64 & 64 & 32\\ 32 & 64 & 32 & 32\\ 32 & 32 & 32 & 32\\ \end{tabular} & 128 & \cellcolor{red!25} Not winnable\\
  \hline
  $4 \times 4$ & \begin{tabular}{ c c c c } 64 & 64 & 64 & 32\\ 64 & 64 & 64 & 32\\ 64 & 32 & 32 & 32\\ 32 & 32 & 32 & 32\\ \end{tabular} & 128 & \cellcolor{red!25} Not winnable\\
  \hline
\end{tabular}

\newpage

\begin{tabular}{ | c | c | c | c | }
  \hline
  Field size & Max allowed tiles during play & Goal & Result\\
  \hline\hline
  $4 \times 4$ & \begin{tabular}{ c c c c } 32 & 64 & 64 & 32\\ 32 & 64 & 32 & 32\\ 32 & 64 & 32 & 32\\ 32 & 32 & 32 & 32\\ \end{tabular} & 128 & \cellcolor{red!25} Not winnable\\
  \hline
  $4 \times 4$ & \begin{tabular}{ c c c c } 32 & 32 & 32 & 32\\ 32 & 64 & 64 & 32\\ 32 & 64 & 64 & 32\\ 32 & 32 & 32 & 32\\ \end{tabular} & 128 & \cellcolor{green!25} Winnable\\
  \hline
  $4 \times 4$ & \begin{tabular}{ c c c c } 64 & 64 & 64 & 64\\ 64 & 128 & 128 & 64\\ 64 & 128 & 128 & 64\\ 64 & 64 & 64 & 64\\ \end{tabular} & 256 & \cellcolor{red!25} Not winnable\\
  \hline\hline
  $2 \times 2 \times 2$ & \begin{tabular}{ c p{40pt} c c } 16 & 16 & 16 & 16\\ 16 & 16 & 16 & 16\\ \end{tabular} & 32 & \cellcolor{green!25} Winnable\\
  \hline
  $2 \times 2 \times 2$ & \begin{tabular}{ c p{40pt} c c } 32 & 32 & 32 & 32\\ 32 & 32 & 32 & 32\\ \end{tabular} & 64 & \cellcolor{red!25} Not winnable\\
  \hline\hline
  $2 \times 2 \times 3$ & \begin{tabular}{ c c p{40pt} c c c } 128 & 128 & 128 & 128 & 128 & 128\\ 128 & 128 & 128 & 128 & 128 & 128\\ \end{tabular} & 256 & \cellcolor{green!25} Winnable\\
  \hline
  $2 \times 2 \times 3$ & \begin{tabular}{ c c p{40pt} c c c } 256 & 256 & 256 & 256 & 256 & 256\\ 256 & 256 & 256 & 256 & 256 & 256\\ \end{tabular} & 512 & \cellcolor{red!25} Not winnable\\
  \hline
\end{tabular}

\subsection{Approximate lower bounds on probabilities}

These were computed using floating-point numbers and therefore cannot be considered proved due to possible rounding errors. In these computations, the goal is changed to that provided by Lemma~\ref{lemma:the-lemma}.\\

\begin{tabular}{ | c | p{120pt} | c | p{50pt} | p{60pt} | }
  \hline
  Field size & Max allowed tiles during play & Goal & Computation data type & Approximate lower bound on probability\\
  \hline\hline
  $3 \times 4$ & \begin{tabular}{ c c c c } 256 & 128 & 64 & 64\\ 64 & 64 & 64 & 64\\ 64 & 64 & 64 & 64\\ \end{tabular} & 512 & \cellcolor{yellow!25} \verb|float32| & 0.999998\\
  \hline
  $3 \times 4$ & \begin{tabular}{ c c c c } 256 & 128 & 16 & 16\\ 64 & 64 & 32 & 16\\ 16 & 16 & 16 & 16\\ \end{tabular} & 512 & \cellcolor{yellow!25} \verb|float32| & 0.999966\\
  \hline
  $3 \times 4$ & \begin{tabular}{ c c c c } 512 & 256 & 128 & 128\\ 128 & 128 & 128 & 128\\ 128 & 128 & 128 & 128\\ \end{tabular} & 1024 & \cellcolor{yellow!25} \verb|float32| & 0.999544\\
  \hline
  $3 \times 4$ & \begin{tabular}{ c c c c } 512 & 256 & 32 & 16\\ 128 & 128 & 64 & 16\\ 16 & 16 & 16 & 16\\ \end{tabular} & 1024 & \cellcolor{yellow!25} \verb|float32| & 0.998857\\
  \hline
  $3 \times 4$ & \begin{tabular}{ c c c c } 1024 & 512 & 128 & 128\\ 256 & 256 & 128 & 128\\ 128 & 128 & 128 & 128\\ \end{tabular} & 2048 & \cellcolor{yellow!25} \verb|float32| & 0.957837\\
  \hline
\end{tabular}

\newpage

\begin{tabular}{ | c | p{120pt} | c | p{50pt} | p{60pt} | }
  \hline
  Field size & Max allowed tiles during play & Goal & Computation data type & Approximate lower bound on probability\\
  \hline\hline
  $3 \times 4$ & \begin{tabular}{ c c c c } 1024 & 512 & 64 & 64\\ 256 & 256 & 128 & 64\\ 64 & 64 & 64 & 64\\ \end{tabular} & 2048 & \cellcolor{yellow!25} \verb|float32| & 0.955559\\
  \hline
  $3 \times 4$ & \begin{tabular}{ c c c c } 1024 & 512 & 64 & 16\\ 256 & 256 & 128 & 16\\ 16 & 16 & 32 & 16\\ \end{tabular} & 2048 & \cellcolor{yellow!25} \verb|float32| & 0.910929\\
  \hline
  $3 \times 4$ & \begin{tabular}{ c c c c } 2048 & 1024 & 128 & 128\\ 512 & 512 & 256 & 128\\ 128 & 128 & 128 & 128\\ \end{tabular} & 4096 & \cellcolor{yellow!25} \verb|float32| & 0.447708\\
  \hline
  $3 \times 4$ & \begin{tabular}{ c c c c } 2048 & 1024 & 128 & 128\\ 512 & 512 & 256 & 128\\ 128 & 128 & 128 & 128\\ \end{tabular} & 4096 & \cellcolor{red!25} \verb|float16| & 0.432\\
  \hline
  $3 \times 4$ & \begin{tabular}{ c c c c } 4096 & 2048 & 256 & 256\\ 1024 & 1024 & 512 & 256\\ 256 & 256 & 256 & 256\\ \end{tabular} & 8192 & \cellcolor{yellow!25} \verb|float32| & 0.000140\\
  \hline\hline
  $3 \times 5$ & \begin{tabular}{ c c c c c } 256 & 128 & 32 & 32 & 32\\ 64 & 64 & 32 & 32 & 32\\ 32 & 32 & 32 & 32 & 32\\ \end{tabular} & 512 & \cellcolor{red!25} \verb|float16| & 0.997\\
  \hline
  $3 \times 5$ & \begin{tabular}{ c c c c c } 256 & 128 & 64 & 32 & 32\\ 64 & 64 & 32 & 32 & 32\\ 64 & 32 & 32 & 32 & 32\\ \end{tabular} & 512 & \cellcolor{red!25} \verb|float16| & 0.997\\
  \hline\hline
  $4 \times 4$ & \begin{tabular}{ c c c c } 1024 & 512 & 64 & 16\\ 256 & 256 & 128 & 16\\ 16 & 16 & 32 & 16\\ 16 & 16 & 16 & 16\\ \end{tabular} & 2048 & \cellcolor{red!25} \verb|float16| & 0.997\\
  \hline
\end{tabular}

\subsection{Exact lower bounds on probabilities}

The integer type used to store exact lower bounds on probability was \verb|uint32_t|.\\

\begin{tabular}{ | c | p{120pt} | c | p{120pt} | }
  \hline
  Field size & Max allowed tiles during play & Goal & Rigorous (proved in integer arithmetic) lower bound on probability\\
  \hline\hline
  $3 \times 4$ & \begin{tabular}{ c c c c } 256 & 128 & 64 & 64\\ 64 & 64 & 64 & 64\\ 64 & 64 & 64 & 64\\ \end{tabular} & 512 & 0.99999617\\
  \hline
  $3 \times 4$ & \begin{tabular}{ c c c c } 512 & 256 & 128 & 128\\ 128 & 128 & 128 & 128\\ 128 & 128 & 128 & 128\\ \end{tabular} & 1024 & 0.99954025\\
  \hline
  $3 \times 4$ & \begin{tabular}{ c c c c } 1024 & 512 & 256 & 256\\ 256 & 256 & 256 & 256\\ 256 & 256 & 256 & 256\\ \end{tabular} & 2048 & 0.95793830\\
  \hline
\end{tabular}

\newpage

\begin{tabular}{ | c | p{120pt} | c | p{120pt} | }
  \hline
  Field size & Max allowed tiles during play & Goal & Rigorous (proved in integer arithmetic) lower bound on probability\\
  \hline\hline
  $3 \times 4$ & \begin{tabular}{ c c c c } 2048 & 1024 & 512 & 512\\ 512 & 512 & 512 & 512\\ 512 & 512 & 512 & 512\\ \end{tabular} & 4096 & 0.45631551\\
  \hline
  $3 \times 4$ & \begin{tabular}{ c c c c c } 4096 & 2048 & 256 & 256\\ 1024 & 1024 & 512 & 256\\ 256 & 256 & 256 & 256\\ \end{tabular} & 8192 & 0.00012353\\
  \hline\hline
  $4 \times 4$ & \begin{tabular}{ c c c c } 1024 & 512 & 64 & 16\\ 256 & 256 & 128 & 16\\ 16 & 16 & 32 & 16\\ \end{tabular} & 2048 & 0.99969486\\
  \hline
\end{tabular}

\section{Proof of Lemma~\ref{lemma:the-lemma}}

We start by swiping left, and then if the first two rows contain no other tiles except our four, we win with right-up-right, otherwise, we swipe up, then up, then up, and continue swiping up as long as possible, then we are in a situation where the number of tiles in the right half of the first row is at least the same as the number of tiles in the right half of the second row. There are three cases.

\begin{figure}[!h]
  \centering
  \begin{subfigure}{.5\textwidth}
    \centering
    \includegraphics[width=150pt]{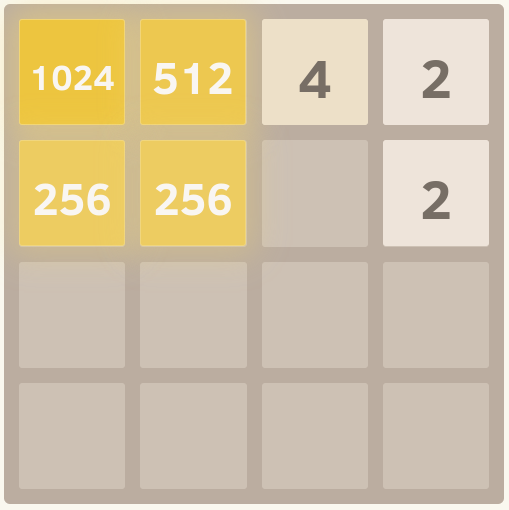}
    \caption{Example of a first case position}
    \label{fig:first-case}
  \end{subfigure}%
  \begin{subfigure}{.5\textwidth}
    \centering
    \includegraphics[width=150pt]{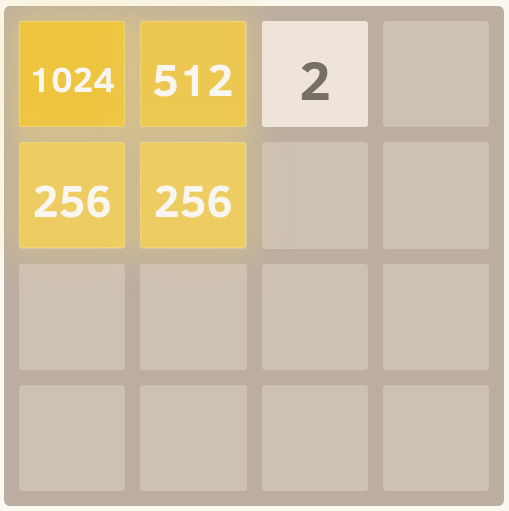}
    \caption{Example of a third case position}
    \label{fig:third-case}
  \end{subfigure}
  \caption{}
\end{figure}

If the first row is filled by now, we have a situation like on Figure~\ref{fig:first-case}, we win with right-left-right-left-\dots and eventually up to merge two 512s.

If the first row has exactly three tiles, and the second row has exactly three tiles, we win with right-up-right.

Finally, if the first row has exactly three tiles, and the second row has exactly two tiles, we are in a situation like on figure~\ref{fig:third-case}. In this case we swipe right, then as many times as we can we swipe left, and if the first row is filled by now, we win with right-left-right-left-\dots-up, otherwise we are in the same situation (except for having 512 instead of two 256s, but it doesn't matter), but \textit{the sum of all tiles in the third and fourth rows has increased}, we repeat the process, notice that the sum can't increase indefinitely.
\hspace*\fill $\blacksquare$

\end{document}